\documentstyle[11pt]{article}
\begin{document}
\begin{center}
{\LARGE{\bf Energy and momentum associated with a Static Axially
Symmetric Vacuum Space-Time}}\\[2em]
\large{\bf{Ragab M. Gad}\footnote{Email Address: ragab2gad@hotmail.com}}\\
\normalsize {Mathematics Department, Faculty of Science,}\\
\normalsize  {Minia University, 61915 El-Minia,  EGYPT.}
\end{center}
\begin{abstract}
We use the Einstein and Papapetrou energy-momentum complexes to
calculate the energy and momentum densities of  Weyl metric as
well as Curzon metric. We show that these two different
definitions of energy-momentum complexes do not provide the same
energy density for Weyl  metric, although they give the same
momentum density. We show that, in the case of Curzon metric,
these two definitions give the same energy only when $R
\rightarrow \infty$. Furthermore, we compare these results with
those obtained using Landau and Lifshitz, Bergmann and M{\o}ller.
\end{abstract}

%%% ----------------------------------------------------------------------
%\maketitle
%%% ----------------------------------------------------------------------

\section{Introduction}
One of the most interesting and intricate problems which remains
unsolved since the outset of the general theory of relativity is
the energy-momentum localization. After Einstein's energy-momentum
complex (E) \cite{E},  used for calculating energy and momentum in
a general relativistic systems,  many physicists, such as, Tolman
(T) \cite{T}, Landau and Lifshitz (LL) \cite{LL}, Papapetrou (P)
\cite{P}, Bergmann (B) \cite{B} and Weinberg (W) \cite{W}
(abbreviated to (ETLLPBW), in the sequel) had given different
definitions for the energy-momentum complexes. These definitions
were restricted to the use of quasi-Cartesian coordinates.
M{\o}ller (M) \cite{E1} introduced a consistent expression which
enables one to evaluate energy and momentum in any coordinate
system. Some interesting results obtained recently lead to the
conclusion that these prescriptions give the same energy
distribution  for a given space-time \cite{V1}-\cite{V8}.
Aguirregabiria, Chamorro and Virbhadra \cite{V9} showed that the
five different\footnote{Virbhadra \cite{V99} has shown that
Tolman's and Einstein's are exactly the same} energy-momentum
complexes (ELLPBW)  give the same result for the energy
distribution for any Kerr-Schild metric. Recently, Virbhadra
\cite{V99} investigated whether or not these definitions (ELLPBW)
lead to the same result for the most general non-static
spherically symmetric metric and found they disagree. He noted
that the energy-momentum complexes (LLPW)  give the same result as
in the Einstein prescription if the calculations are performed in
Kerr-Schild Cartesian coordinates. However, the complexes (ELLPW)
disagree  if computations are done in ``Schwarzschild Cartesian
coordinates \footnote{ Schwarzschild metric in ``Schwarzschild
Cartesian coordinates'' is obtained by transforming this metric
(in usual Schwarzschild coordinates $\{t, r, \theta, \phi\}$)  to
$\{t,x,y,z\}$ using $ x = r \sin\theta \cos\phi, x = r \sin\theta
\sin\phi, z = r \cos\theta $.}.
\par
Some interesting results \cite{Xulu1}-\cite{Gad3} led to the
conclusion that in a given space-time, such as: the
Reissner-Nordst\"{o}rm, the de Sitter-Schwarzschild, the charged
regular metric, the stringy charged black hole and the
G\"{o}del-type space-time, the energy distribution according to
the energy-momentum complex of Einstein is different from of
M{\o}ller. But in some specific case \cite{E1,Xulu1,EV2,V97,V99}
(the Schwarzschild, the Janis-Newman-Winicour metric) have the
same result.

\par
The scope of this paper is to evaluate  the energy  and momentum
densities for the Weyl as well as Curzon metrics using the
Einstein and Papapetrou energy-momentum complexes. Through this
paper we use $G = 1$ and $c = 1$ units and follow the convention
that Latin indices take value from 0 to 3 and Greek indices take
value from 1 to 3.
\par
The general static axially symmetric vacuum solution of Einstein's field equations is given by the
Weyl metric \cite{11}
\begin{equation} \label{1}
ds^2 = e^{2\lambda} dt^2 - e^{2(\nu -\lambda )}(dr^2 + dz^2) - r^2 e^{-2\lambda} d\phi^2
\end{equation}
where
$$
\lambda_{rr} + \lambda_{zz} + r^{-1}\lambda_{r} = 0
$$
and
$$
\nu_{r} = r(\lambda^{2}_{r} - \lambda^{2}_{z}), \qquad \nu_{z} = 2r\lambda_{r}\lambda_{z}.
$$

\par
It is well known that if the calculations are performed in
quasi-Cartesian coordinates,
all the energy-momentum complexes give meaningful results.\\
According to the following transformations
$$
r = \sqrt{x^2 + y^2} , \qquad \phi = \arctan(\frac{y}{x}),
$$
the line element (\ref{1}) written in terms of  quasi-Cartesian
coordinates reads:
$$
ds^2 = e^{2\lambda}dt^2 - \frac{1}{r^2}\big( x^2e^{2(\nu -
\lambda)} + y^2e^{-2\lambda} \big) dx^2 -
\frac{2xy}{r^2}\Big(e^{2(\nu - \lambda)} - e^{-2\lambda}\Big)dxdy-
$$
\begin{equation} \label{2}
  \frac{1}{r^2} \big( y^2e^{2(\nu - \lambda)} + x^2e^{-2\lambda}\big)dy^2  -  e^{2(\nu - \lambda)}dz^2,
\end{equation}
where
$$
x^2\lambda_{xx} + y^2\lambda_{yy} + 2xy\lambda_{xy} +
r^2\lambda_{zz} + x\lambda_{x} + y\lambda_{y} = 0,
$$
$$
x\nu_{x} + y\nu_{y} -(x\lambda_{x} + y\lambda_{y})^2
+r^2\lambda_{z} = 0
$$
and
$$
\nu_{z} = 2\lambda_{z}(x\lambda_{x} + y\lambda_{y}).
$$
\par
 For the above metric the determinant of the metric tensor
and the contravariant components of the tensor are given,
respectively, as follows
\begin{equation}\label{3}
%\begin{split}
\begin{array}{ccc}
det g & = &- e^{4(\nu - \lambda)},\\
g^{00} & = &e^{-2\lambda}, \\
g^{11} & = &- \frac{e^{2\lambda }}{r^2} (y^2 + x^2 e^{-2\nu}),\\
g^{12} & = & \frac{xye^{2\lambda}}{r^2}(1-e^{-2\nu}),\\
g^{22} & = &-\frac{e^{2\lambda}}{r^2}(x^2+y^2e^{-2\nu}),\\
g^{33} & = &- e^{2(\lambda - \nu)}.
%\end{split}
\end{array}
\end{equation}
\section{\bf{Energy-Momentum Complexes}} The conservation laws of
matter plus non-gravitational fields for physical system in the
special theory of relativity are given by
\begin{equation}\label{EM1}
T_{\nu,\mu}^{\mu} \equiv \frac{\partial T_{\nu}^{\mu}}{\partial
x^{\mu}} = 0,
\end{equation}
where $T_{\nu}^{\mu}$ denotes the symmetric energy-momentum tensor
in an inertial frame.
\par
 The generalization of equation (\ref{EM1})
in the theory of general relativity is written as
\begin{equation}\label{EM2}
T_{\nu;\mu}^{\mu} = \frac{1}{\sqrt{-g}}\frac{\partial}{\partial
x^{\mu}}\big(\sqrt{-g}
T_{\nu}^{\mu}\big)-\Gamma_{\nu\lambda}^{\mu}T_{\mu}^{\lambda} = 0,
\end{equation}
where $g$ is the determinant of the metric tensor $g_{\mu\nu}(x)$.
\par
The conservation equation may also be written as
\begin{equation}\label{EM3}
\frac{\partial}{\partial x^{\mu}}\big(\sqrt{-g}T_{\nu}^{\mu}\big)=
\xi_{\nu},
\end{equation}
where
$$
\xi_{\nu} = \sqrt{-g} \Gamma_{\nu\lambda}^{\mu}T_{\mu}^{\lambda}
$$
is a non-tensorial object and it can be written as
\begin{equation}\label{EM4}
\xi_{\nu} = -\frac{\partial}{\partial
x^{\mu}}\big(\sqrt{-g}t^{\mu}_{\nu}\big).
\end{equation}
where $t^{\mu}_{\nu}$ are certain functions of the metric tensor
and its first derivatives.
\par
 Now combining equation (\ref{EM4}) with equation (\ref{EM3}) we
 get the following equation expressing a local conservation
 law:
 \begin{equation}\label{EM5}
 \Theta^{\mu}_{\nu,\mu} = 0,
 \end{equation}
 where
 \begin{equation}\label{EM6}
 \Theta_{\nu}^{\mu} = \sqrt{-g}\big( T_{\nu}^{\mu} +
 t^{\mu}_{\nu}\big)
 \end{equation}
 which is called energy-momentum complex since it is a combination
 of the tensor, $T_{\nu}^{\mu}$, of matter and all non-gravitational fields, and a pseudotensor
 $t^{\mu}_{\nu}$ which describes the energy and momentum of the
 gravitational field\footnote{By introducing a local system of inertia,
 the gravitational part $t^{\mu}_{\nu}$ can always be reduced to
 zero for any given space-time.}.\\
 Equation (\ref{EM6}) can be written as
 \begin{equation}
 \Theta_{\nu}^{\mu} = \chi^{\mu\lambda}_{\nu,\lambda},
 \end{equation}
 where $\chi_{\nu}^{\mu\lambda}$ are called superpotentials and are
 functions of the metric tensor and its first derivatives.

\section{\bf{Energy-momentum Complex in Einstein's Prescription}}
The energy-momentum complex as defined by Einstein \cite{E} is
given by
\begin{equation} \label{3.1}
\theta^{k}_{i} = \frac{1}{16\pi}H^{kl}_{\, \,\, i,l},
\end{equation}
where the Einstein's superpotential $H^{kl}_{\, \,\, i}$ is of the
form
\begin{equation} \label{3.2}
H^{kl}_{\, \,\, i} = - H^{lk}_{\, \,\, i} = \frac{g_{in}}{\sqrt{-
g}} \big[ - g\big( g^{kn}g^{lm} - g^{ln}g^{km}\big)\big]_{,m}.
\end{equation}
$\theta^{0}_{0}$ and $\theta^{0}_{\alpha}$ are the energy and
momentum density components, respectively.\\
The energy-momentum complex $\theta^k_i$ satisfies the local
conservation law
$$
\frac{\partial\theta^k_i}{\partial x^k} = 0
$$
\par
In order to evaluate the energy and momentum densities in
Einstein's prescription associated with the Weyl metric, we
evaluate the non-zero components of $H^{kl}_{\, \,\, i}$

\begin{equation}\label{3.5}
%\begin{split}
\begin{array}{ccc}
H^{01}_{\, \,\, 0}& =
\frac{1}{r^2}\Big[xye^{2\nu}(4\nu_{y}-4\lambda_{y}) -
xy(2\nu_{y}-4\lambda_{y}) - y^2e^{2\nu}(4\nu_{x}-4\lambda_{x})-
\\&
x^2(2\nu_{x}-4\lambda_{x}) + x(e^{2\nu}-1)\big]\\
H^{02}_{\, \,\, 0} &=
\frac{1}{r^2}\Big[xye^{2\nu}(4\nu_{x}-4\lambda_{x}) -
xy(2\nu_{x}-4\lambda_{x}) - x^2e^{2\nu}(4\nu_{y}-4\lambda_{y})-
\\&
y^2(2\nu_{y}-4\lambda_{y}) + y(e^{2\nu}-1)\big]\\
 H^{03}_{\, \,\, 0} & = 4\lambda_{z} -
2\nu_{z}.
%\end{split}
\end{array}
\end{equation}
Using these components  in equation (\ref{3.1}), we get the energy
and momentum densities as following
%\begin{equation}\label{3.6}
%\begin{split}
%\begin{align*}
$$
\begin{array}{ccc}
 \theta^{0}_{0}& =
\frac{1}{16\pi r^2}\Big[(2\nu_{x}-4\lambda_{x})(xe^{2\nu} +
2ye^{2\nu}(x\nu_{y}-y\nu_{x})) +(2\nu_{y}-4\lambda_{y})(ye^{2\nu}
+
\\& 2xe^{2\nu}(y\nu_{x}-x\nu_{y}))+
4e^{2\nu}(y\nu_{y}+x\nu_{x})-4e^{2\nu}(x\nu_{y}-y\nu_{x})^2
-4xy\nu_{xy} -\\
&y^2 e^{2\nu}(4\nu_{xx}-4\lambda_{xx}) -
x^2e^{2\nu}(4\nu_{yy}-4\lambda_{yy})- 2x^2 \nu_{xx}
-2y^2\nu_{yy}-\\
&2(x\nu_{x}+y\nu_{y}) + 2xye^{2\nu}(4\nu_{xy}-4\lambda_{xy})
-2r^2\nu_{zz}\Big],
%\end{split}
%\end{equation}
%\end{align*}
\end{array}
$$
$$
\theta_{\alpha}^{0} = 0.
$$
The momentum components are vanishing everywhere.
\par
We now restrict our selves to the particular solutions of  Curzon
metric \cite{Curzon} obtained by setting
$$
\lambda = - \frac{m}{R} \qquad and \quad \nu = -\frac{m^2
r^2}{2R^4}, \qquad R = \sqrt{r^2 + z^2}
$$
in equation (\ref{1}). \\
For this solution it is found from equation (\ref{3.5}) that the
non-zero components of $H^{kl}_{\, \,\, i}$ take the form
\begin{equation}\label{3.8}
%\begin{split}
\begin{array}{ccc}
H^{01}_{\, \,\, 0}& = &x\Big[\frac{2m^2}{R^4} -
\frac{4m^2r^2}{R^6}+
\frac{4m}{R^3} + \frac{1}{r^2}(e^{2\nu} - 1)\Big]\\
H^{02}_{\, \,\, 0} &= &y\Big[\frac{2m^2}{R^4} -
\frac{4m^2r^2}{R^6}+
\frac{4m}{R^3} + \frac{1}{r^2}(e^{2\nu} - 1)\Big]\\
 H^{03}_{\, \,\, 0} & =& \frac{4mz}{R^3}\big[ 1 - \frac{mr^2}{R^3}\big].
%\end{split}
\end{array}
\end{equation}
Using these components the energy and momentum densities for the
Curzon solution become
\begin{equation}\label{3.9}
\theta^{0}_{0} = \frac{1}{16\pi}\Big[-\frac{4m^2r^2}{R^6} +
\frac{4m^2}{R^4} + 2e^{2\nu}\big( -\frac{m^2}{R^4} +
\frac{2m^2r^2}{R^6}\big)\Big],
\end{equation}
\begin{equation}\label{3.10}
\theta_{\alpha}^{0} = 0.
\end{equation}
The momentum components are vanishing everywhere.
\par

\section{\bf{The Energy-Momentum Complex of Papapetrou}}
The symmetric energy-momentum complex of Papapetrou \cite{P} is
given by
\begin{equation}\label{6.1}
\Omega^{ij} = \frac{1}{16\pi} \Upsilon^{ijkl}_{\quad ,kl},
\end{equation}
where
\begin{equation}\label{6.2}
\Upsilon^{ijkl} = \sqrt{- g}\big( g^{ij}\eta^{kl} -
g^{ik}\eta^{jl} + g^{kl}\eta^{ij} - g^{jl}\eta^{ik}\big),
\end{equation}
and $\eta^{ik}$ is the Minkowski metric with signature $-2$.\\
$\Omega^{00}$ and $\Omega^{\alpha 0}$ are the energy and momentum
density components. In order to calculate the energy and momentum
density components for Weyl metric, using the symmetric energy
momentum complexes of Papapetrou, we require the following
non-vanishing components of $\Upsilon^{ijkl}$
\begin{equation}\label{6.5}
%\begin{split}
\begin{array}{ccc}
\Upsilon^{0011} & =& - \big( e^{(2\nu - 4\lambda)} + \frac{y^2e^{2\nu}+ x^2}{r^2}\big),\\
\Upsilon^{0012} & = &\frac{xy}{r^2}\big(e^{2\nu}-1\big)\\
\Upsilon^{0022} & = &- \big( e^{(2\nu - 4\lambda)} + \frac{x^2e^{2\nu}+ y^2}{r^2}\big),\\
\Upsilon^{0033} & = &- \big( e^{(2\nu - 4\lambda)} + 1 \big).
%\end{split}
\end{array}
\end{equation}
Using these components in (\ref{6.1}), we get the energy and
momentum densities in the following form
$$
\Omega^{00} = \frac{1}{16\pi}\Big[-e^{2\nu -
4\lambda}\big((2\nu_{x} - 4\lambda_{x})^2 + 2\nu_{xx} -
4\lambda_{xx} + (2\nu_{y} - 4\lambda_{y})^2 + 2\nu_{yy} -
4\lambda_{yy} +
$$
$$
(2\nu_{z} - 4\lambda_{z})^2 + 2\nu_{zz} - 4\lambda_{zz}\big) -
\frac{4e^{2\nu}}{r^2}(y\nu_{x} - x\nu_{y})^2 -
\frac{2y^2e^{2\nu}}{r^2}\nu_{xx} -
\frac{2x^2e^{2\nu}}{r^2}\nu_{yy} +
$$
\begin{equation}\label{6.6}
\frac{4ye^{2\nu}}{r^2}\nu_{y} + \frac{4xe^{2\nu}}{r^2}\nu_{x} +
\frac{4xye^{2\nu}}{r^2}\nu_{xy}\Big].
\end{equation}
\begin{equation}\label{6.7}
\Omega^{\alpha 0} = 0.
\end{equation}
For the Curzon solution the energy and momentum densities become
$$
\Omega^{00} = \frac{1}{16\pi}\Big[-e^{2\nu - 4\lambda}\big(
\frac{4m^4r^2}{R^8} + \frac{12m^2}{R^4} - \frac{16m^3r^2}{R^7} +
\frac{4m^2}{R^6}\Big) +
$$
\begin{equation}\label{6.8}
2e^{2\nu}\Big(\frac{2m^2r^2}{R^6}-\frac{m^2}{R^4} \Big)\Big]
\end{equation}
\begin{equation}\label{6.9}
\Omega^{\alpha 0} = 0.
\end{equation}
\par
 In the following table we summarize our results
obtained  (see, \cite{Gad4}) of the energy and momentum densities
for Curzon metric, using  Landau and Lifshitz,  Bergmann and
M{\o}ller.
\newpage

\begin{table}
  \centering
  \begin{tabular}{|c|c|c|}
  %\begin{tabular}{|t|l|}
\hline
    {\bf{Prescription}}& {\bf{Energy density}} & {\bf{Momentum density}} \\
     \hline
    Landau and Lifshitz& $L^{00} = \frac{1}{8\pi}e^{4\nu - 4\lambda}
    \Big[-\frac{2m^2}{R^4}+\frac{4m^2r^2}{R^6}-\frac{2m}{R^3}+$ &\\
&$e^{-2\nu}\Big( -\frac{5m^2}{R^4} - \frac{4m^2r^2}{R^6}-
\frac{2m^4r^2}{R^8}+\frac{8m^3r^2}{R^7}+\frac{2m}{R^3}\Big)\Big]$    & $L^{\alpha 0} =0$\\
    \hline
    Bergmann& $B^{00}= \frac{me^{-2\lambda}}{8\pi R^{3}}\Big[ -\frac{2m}{R} +
    \frac{2m^{2}r^2}{R^4} -(e^{2\nu} -1)-$&\\
    & $\frac{2mr^2}{R^3}- \frac{me^{2\nu}}{R} + \frac{2mr^2e^{2\nu}}{R^3}\Big]$&$B^{\alpha 0}=0$\\
    \hline
     M{\o}ller & $\Im^{0}_{0} = \frac{m}{4\pi R^3}\Big[2(r+z) - \frac{3}{R^2}(r^3 +
z^3)\Big]$ & $\Im_{\alpha}^{0} = 0$.\\
\hline
  \end{tabular}
  \vspace{2mm}
  \caption{\sf{The energy and momentum densities, using (LLBM),
   for the Curzon metric}}
\end{table}

\section*{\bf{Discussion}}

The problem of energy-momentum localization associated with much
debate.
 Misner et al. \cite{MTW} argued that the energy is localizable only for
spherical systems. There are other  opinions which contradict
their viewpoint. Cooperstock and Sarracino \cite{CS} argued that
if the energy localization is meaningful for spherical systems
then it is meaningful for all systems. Bondi \cite{Bo} expressed
that a non-localizable form of energy is inadmissible in
relativity and its location can in principle be found.
\par
 Using different definitions of energy-momentum complex, several
authors  studied the energy distribution for a given space-time.
Most of them restricted their intention to the static and
non-static spherically symmetric space-times. Rosen and Virbhadra
\cite{RV} calculated the energy and momentum densities of
non-static cylindrically symmetric empty space-time.
\par
 In this paper, we
calculated the energy and momentum density components for Weyl
metric as well as Curzon metric using Einstein and Papapetrou
prescriptions.

We found that for both considered Weyl and Curzon metrics, the
Einstein and Papapetrou give exactly the same momentum density but
do not provide the same energy density, except only at $R
\rightarrow \infty$, in the case of Curzon metric, where the
energy density tends to zero.
\par
Furthermore, we have made a comparison of our results with those
calculated \cite{Gad4} using (LLBM) prescriptions. We obtained
that the five prescriptions (ELLPBM) give the same result
regarding the momentum density associated with Weyl as well as
Curzon metrics. Concerning the energy density associated with both
two metrics under consideration, we found that these prescriptions
(ELLPBM) do not give the same result except when $R \rightarrow
\infty$, in the case of Curzon metric, where the energy in the all
prescriptions (ELLPBM) tends to zero.\\
Finally, in the case of Curzon metric we see that the energy in
the all prescriptions (ELLPBM) diverge at the singularity ($R =
0$).

\end{document}